\documentclass[info,overfull]{epl}%
\usepackage{graphicx}
\usepackage{amsmath}
\usepackage{amsfonts}
\usepackage{amssymb}%
\institute{
\inst{1}
Laboratoire de Physique de la Mati\`{e}re Condens\'{e}e, UMR 7125 du CNRS,
Coll\`{e}ge de France, 75231 Paris Cedex 05, France\\
\inst{2}
Department of Physics, Graduate School of Humanities and Sciences, Ochanomizu
University, 2--1--1, Otsuka, Bunkyo-ku, Tokyo 112-8610, Japan\\
\inst{3}
Institut de Recherche sur les Ph\'{e}nom\`{e}nes Hors Equilibre,
UMR 6594 du CNRS, BP 146, 13384 Marseille Cedex, France
}
\begin{document}

\title{Water spring: a model for bouncing drops}
\author{K. Okumura,\inst{1,2} F. Chevy,\inst{1} D. Richard,\inst{1} D.
Qu\'{e}r\'{e}\inst{1} and C. Clanet\inst{3}}
\date{\today    }
\maketitle

\begin{abstract}
It has been shown that a water drop can bounce persistently, when thrown on a
super-hydrophobic substrate. We present here scaling arguments which allow us
to predict the maximal deformation and the contact time of the drop. This
approach is completed by a model describing the flow inside the drop, and by
original experimental data.

\end{abstract}

\shortauthor{K. Okumura \etal}

\section{Introduction}

A \emph{liquid ball} is a drop which remains quasi-spherical when brought into
contact with a solid surface.\ Different examples of such objects have
recently been described: let us quote \textit{pearl drops}, which result from
the extreme hydrophobicity of the solid \cite{Kao,KT,JB,text}, \textit{liquid
marbles}, achieved by texturing the surface of the liquid \cite{LM,6,7}, and
(more classically) \textit{Leidenfrost drops}, obtained by putting a small
amount of volatile liquid on a very hot plate \cite{FR}. In such cases, quick
transportation of tiny amounts of liquid becomes possible without any leak,
which can be of great interest in microfluidics applications. At the same time
these systems realize pure\ capillary "devices," and are worth being studied
for their original properties.

For example, a liquid ball (with a typical diameter 1 mm) impinging onto a
solid substrate bounces off, as if it were a tennis ball hitting the ground
\cite{leaf,DD}. The rebounds are persistent (the restitution coefficient can
be very high, of the order of 0.9), and observed in a large window of impact
velocity.\ If the velocity is too small (typically smaller than a few
centimeters per second), the drop gets stuck on the substrate.\ If too large
(above around 1 m/s), the drop endures extreme deformation during the shock,
and finally breaks into several pieces. In between, a more detailed analysis
of the shock itself shows that the contact time of the drop with the substrate
does not depend on the impact velocity $V$ over a large interval of velocity
\cite{DR}.

In this note, we discuss the maximal deformation of this sort of "water
spring," and also the value of the contact time during the shock, in
particular in the limit of elastic impact (small impact velocity). These
predictions are compared with original experimental data.

\begin{figure}[ptb]
\includegraphics[scale=0.5]{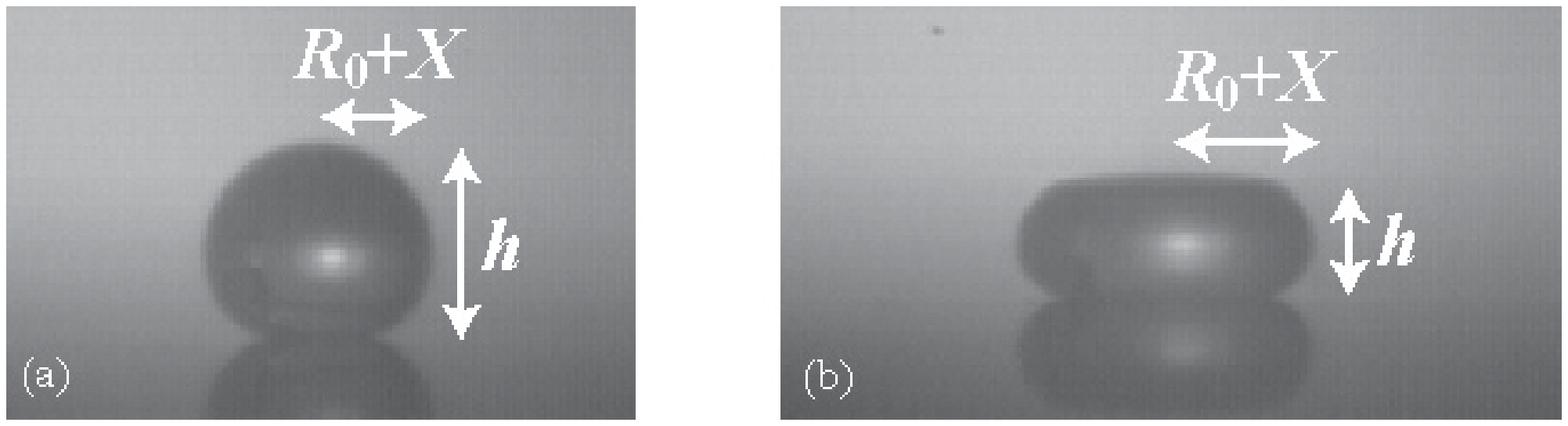}\caption{Maximal deformation of a liquid
ball impinging a solid at a small velocity (a: $R_{0}\sim0.4$ mm, $V=0.094$
m/s, $W\sim0.05$); or at a larger velocity (b: $R_{0}\sim0.4$ mm, $V=0.47$
m/s, $W\sim1.$). }%
\label{f1}%
\end{figure}

We shall mainly focus on the limit of small deformations where a drop of
initial radius $R_{0}$ bounces back with a high restitution coefficient of the
order of 0.9. This corresponds to kinetic energies smaller than surface
energies, \textit{i.e.} to small Weber numbers, where the latter quantity is
defined as
\begin{equation}
W=\rho R_{0}V^{2}/\gamma.\label{weber}%
\end{equation}
This condition is achieved for small impact velocities, \textit{i.e.} smaller
than $V_{W}$:
\begin{equation}
V_{W}=\sqrt{\frac{\gamma}{\rho R_{0}}}.\label{vw}%
\end{equation}
For a millimetric water drop ($\gamma\sim72$ mN/m$,\rho=10^{3}$ kg/m$^{3}$),
$V_{W}$ is of the order of 1 m/s. As seen in Fig. \ref{f1}, the drop at its
maximal deformation looks like a flattened sphere for small impact velocities
($V<V_{W}$), or a pancake for larger velocities ($V\sim V_{W}$).

\section{Scaling equations for a water spring ($V\ll V_{W}$)}

We start from the Euler's equation:%
\begin{equation}
\rho\frac{Dv}{Dt}=-\nabla p+\rho g, \label{Euler}%
\end{equation}
neglecting the effect of viscosity thanks to the high restitution coefficient.
We consider the limit $V\ll V_{W}$, in which the drop is deformed by a
quantity $X$ much smaller than the initial radius $R_{0}$ (see Fig.
\ref{f1}.a); the characteristic time of deformation scales as $X/V$. On the
other hand, the Laplace pressure gradient scales as $\gamma X/R_{0}^{3}$ (for
example, assuming an ellipsoidal form for the drop, the pressure jump at the
equator or the apex scales as $\Delta p\sim\gamma/R_{0}(1\pm X/R_{0})$ and
changes over the length $R_{0}$). Thus, Eq. (\ref{Euler}) can be dimensionally
written as%
\begin{equation}
\rho V^{2}R_{0}^{3}\sim\gamma X^{2}-\rho gR_{0}^{3}X. \label{ec2}%
\end{equation}
This equation also expresses the transfer of kinetic energy into the surface
and gravitational terms associated with the drop deformation. If the velocity
is zero, the drop is deformed by gravity by a quantity $\delta$, as first
discussed by Mahadevan and Pomeau \cite{MP}, which reads
\begin{equation}
\delta\sim R_{0}^{3}/\kappa^{-2}, \label{delta}%
\end{equation}
by use of the capillary length $\kappa^{-1}=\sqrt{\gamma/\left(  \rho
g\right)  }$, which is about 3 mm for water.

We first consider the case where capillarity dominates gravity ($X\gg\delta$).
Then, we find from Eq. (\ref{ec2}) that the maximal deformation scales as the
velocity:%
\begin{equation}
X\sim\left(  \rho R_{0}^{3}/\gamma\right)  ^{1/2}V. \label{3}%
\end{equation}
The contact time should be of the order of $X/V,$ and thus can be written as%
\begin{equation}
\tau\sim\left(  \rho R_{0}^{3}/\gamma\right)  ^{1/2}. \label{4}%
\end{equation}
Eq. (\ref{4}), which yields a $V$-independent contact time of about one
millisecond, indeed corresponds to the plateau observed experimentally
\cite{DR}.

Eq. (\ref{3}) allows us to make clear the limit of this regime. Since $X$ must
be larger than $\delta$, we find that the velocity must be larger than a
characteristic velocity,%
\begin{equation}
V_{c}\sim\left(  gR_{0}^{3}\kappa^{2}\right)  ^{1/2},
\end{equation}
which is about a few cm/s. If $V$ approaches $V_{c}$, gravity cannot be
neglected any more; we recast Eq. (\ref{ec2}) as%
\begin{equation}
\rho V^{2}R_{0}^{3}+\gamma\delta^{2}\sim\gamma\left(  X-\delta\right)
^{2}\label{spring}%
\end{equation}
which expresses the energy conservation of an \textit{imaginary} spring-mass
system of initial velocity $V$ and initial elongation $\delta$.
\begin{figure}[!]
\includegraphics[scale=0.6]{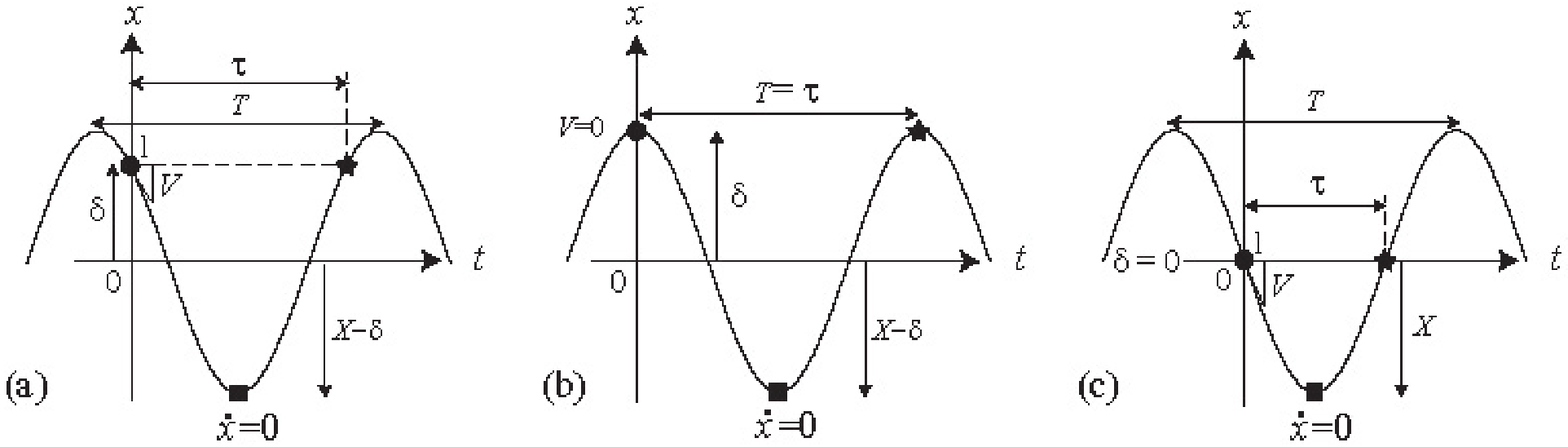}
\caption
{Analogy with a spring system: elongation $x$ is plotted as a function of
time $t$. The circle ($t=0$) corresponds to the
moment of impact and the square to the maximal deformation.
The star thus indicates the moment of rebound. (a) general case. (b)
small velocity limit. (c) large velocity limit.}
\label{f2}
\end{figure}%

Fig. \ref{f2} may help to have an intuitive picture for Eq. (\ref{spring}).
The elongation $x$ of the imaginary spring is plotted as a function of time
$t$, setting $t=0$ at the moment of impact. The left-hand side of Eq.
(\ref{spring}) is represented by a circle ($x=\delta,$ $\dot{x}=V$ at $t=0$)
while the right-hand side by a square ($\dot{x}=0$ at $t=\tau/2$). The moment
of taking off is marked by a star, and $\tau$ is compared in the plot with the
period of the oscillator $T\sim\left(  \rho R_{0}^{3}/\gamma\right)  ^{1/2}$.

Thus, we can graphically deduce the contact time $\tau$. In the small velocity
limit $V\ll V_{c}$ ($\rho V^{2}R_{0}^{3}\ll\gamma\delta^{2}$), we see in Fig.
\ref{f2}.b that the contact time $\tau$ is equal to the vibration period $T.$
In contrast, in the interval commented above ($V_{c}\ll V\ll V_{W}$) and with
increase in $V$, $\tau$ decreases from $T$ and approaches the plateau value
$T/2$. The complete variation of the contact time $\tau$ as a function of the
impact velocity $V$ will be calculated more precisely in the next section.
Note finally that, as stated above, the maximal deformation, in the limit of
extremely small impact velocities, tends towards the constant $\delta$, as
seen from Eq. (\ref{spring}): we logically recover the static deformation.

\section{Local model}

The previous scaling arguments can be completed by considering local flows
during the impact. We start from the incompressibility condition ($\nabla
\cdot\mathbf{v}=0$), which reduces to%
\begin{equation}
\frac{1}{r}\frac{\partial rv_{r}}{\partial r}+\frac{\partial v_{z}}{\partial
z}=0,
\end{equation}
in the cylindrical coordinate system ($r,\theta,z$) with $v_{\theta}=0$.
Looking for a solution of the type $v_{r}=v_{r}(r)$ and $v_{z}=v_{z}(z)$
satisfying appropriate boundary conditions, and introducing the equatorial
radius $R$ (see Fig. \ref{f1}.b), we find%
\begin{equation}
v_{r}=\frac{\dot{R}}{R}r,v_{z}=-2\frac{\dot{R}}{R}z, \label{vloc}%
\end{equation}
setting $z=0$ at the substrate surface. The velocity potential $\phi$ defined
by $\mathbf{v=}\nabla\phi$ can thus be written as
\begin{equation}
\phi=\frac{\dot{R}}{R}\left(  \frac{r^{2}}{2}-z^{2}\right)  .
\end{equation}
If the viscosity is neglected, the dynamics of $\phi$ is governed by
Bernoulli's equation:
\begin{equation}
\rho\frac{\partial\phi}{\partial t}+\frac{1}{2}\rho v^{2}+p+\rho
gz=\text{constant.} \label{Bern}%
\end{equation}
Here, the pressure $p$ at the surface of the drop is given by the Laplace
pressure, i.e. $p=\gamma C(r,z)$ where the local curvature is denoted $C$.

\begin{figure}[!]
\twofigures[scale=0.3]{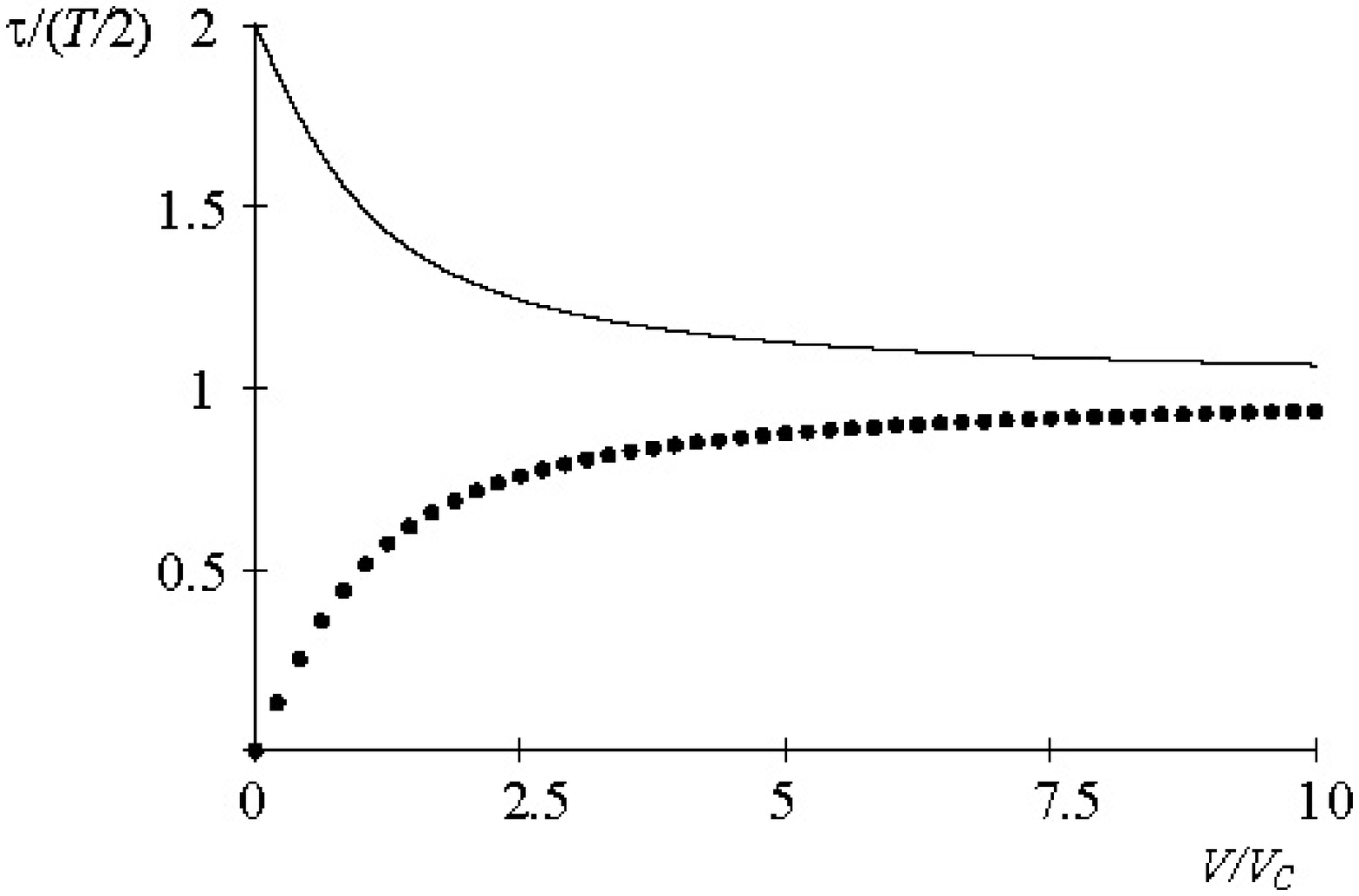}{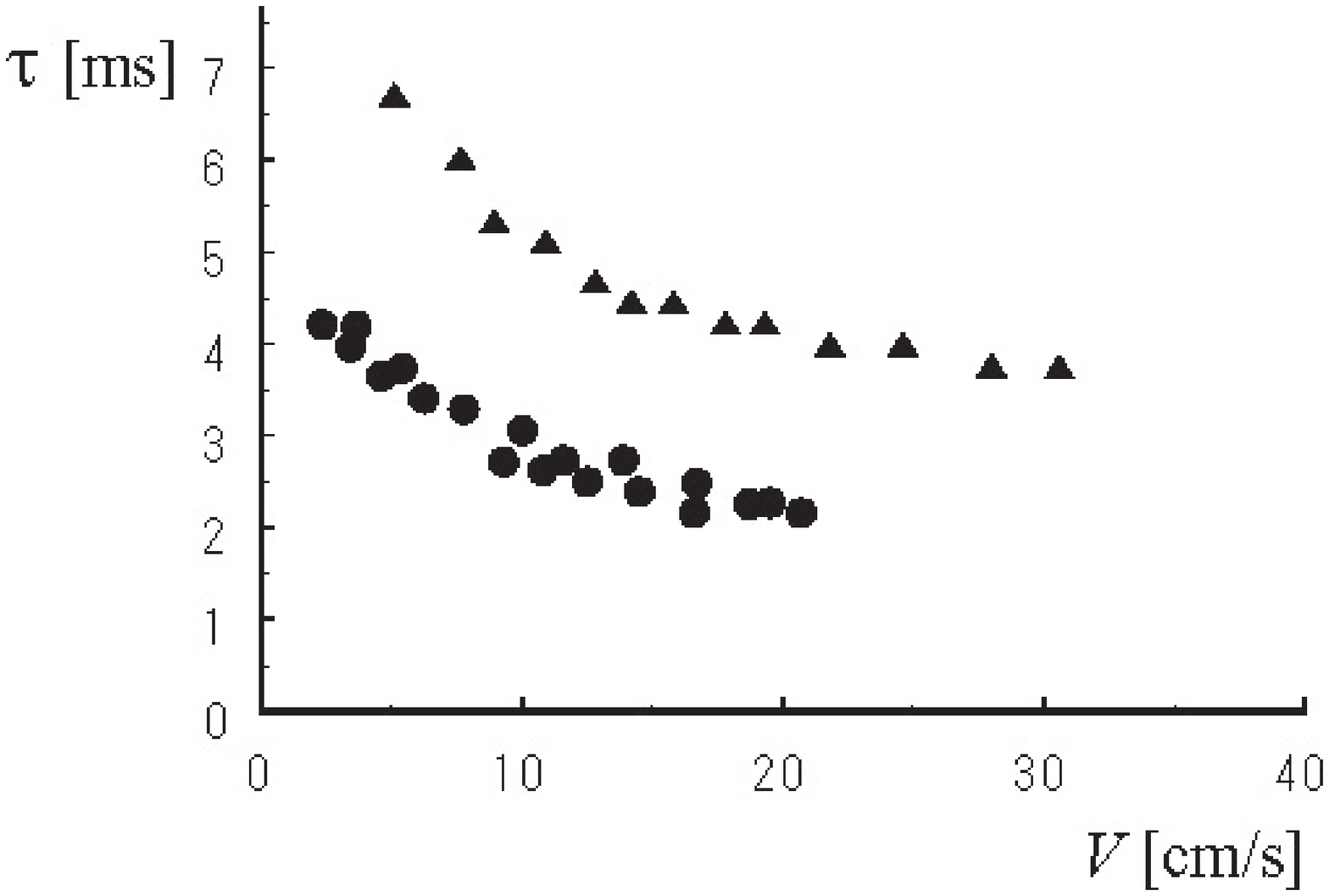}
\caption{Contact time $\tau$ versus the impact
velocity $V$ in the elastic collision limit. The line is calculated from Eq. (\ref
{analytical}).
The dotted line corresponds to the case of inverted gravity (see the text).}
\label{f3}
\caption{Contact time of a water drops bouncing on a super-hydrophobic
substrate as a function of impact velocity. Triangles correspond to a drop
with $R_{0}\sim0.6$ mm, while circles to $R_{0}\sim0.4$
mm.}
\label{f4}
\end{figure}

We evaluate Eq. (\ref{Bern}) at the apex ($r=0,$ $z=h$) and at the equator
($r=R,$ $z\simeq h/2$), in the limit of small deformation ($\ddot{x}\gg\dot
{x}^{2}/R_{0}$) for which $h\simeq2R_{0}$. (Here, $x$ is defined as in Fig.
\ref{f1} but not necessarily at its maximal deformation; the maximum magnitude
of $x$ is $X$.) To estimate the curvature difference $\Delta C=C\left(
0,h\right)  -C\left(  R,h/2\right)  $, we take the value in the static limit
since we are in the regime of small velocities; the condition of constant
pressure inside the drop can be integrated numerically, which gives $\Delta
C=6.8x/R_{0}^{2}$. In this way, we obtain
\begin{equation}
-7\rho R_{0}^{3}\ddot{x}/2+6.8\gamma x-\rho gR_{0}^{3}=0\label{adda1}%
\end{equation}
Writing $\delta=\kappa^{2}R_{0}^{3}/6.8$ and $\omega^{2}=12.8\gamma/(7\rho
R_{0}^{3})$ (which gives as a plateau value for the contact time
$T/2=2.3\sqrt{\rho R_{0}^{3}/\gamma}$), we find a general solution:%
\begin{equation}
x-\delta=x_{0}\cos\left(  \omega t+\varphi\right)  \label{addx}%
\end{equation}
The initial conditions lead to the relations $\cos\varphi=\delta/x_{0}$ and
$\sin\varphi=V/\left(  \omega x_{0}\right)  $, from which we get $x_{0}%
=\sqrt{\delta^{2}+V^{2}/\omega^{2}}$ and $X=\delta+x_{0}$. The time of rebound
$\tau$ should be given by the equation, $\delta=x_{0}\cos(\omega\tau
+\varphi),$ with $\pi\leq\omega\tau+\varphi\leq2\pi$. Hence, we find
$\omega\tau+\varphi=2\pi-\arccos\left(  \delta/x_{0}\right)  $ with
$\varphi=\arccos(\delta/x_{0})$ ($0\leq\varphi\leq\pi$) or%
\begin{equation}
\tau=2\tau_{c}\left(  1-\frac{1}{\pi}\arccos\left(  \frac{\delta}{x_{0}%
}\right)  \right)  \label{analytical}%
\end{equation}
where%
\[
\delta/x_{0}=\left(  1+\left(  V/V_{c}\right)  ^{2}\right)  ^{-1/2}.
\]
The analytical result Eq. (\ref{analytical}) is plotted in Fig. \ref{f3}, and
allows us to recover the predictions made \textit{via} Fig. \ref{f2}: the
contact time decreases by a factor 2 when the impact velocity is increased
from about $V_{c}$ to $V_{W}$.

\section{Experiments}

We measured the contact time of bouncing drops at small impact velocities for
millimetric water droplets hitting a super-hydrophobic substrate. Rebounds
were sequentially recorded with a high-speed camera with a typical sampling
time of 10$^{-4}$ s for a contact time around 10$^{-3}$ s, which gives a
precision higher than 10 \%. The results are plotted in Fig. \ref{f4}\ for two
drop sizes. We indeed observed that the contact time significantly increases
at low impact velocity (in the range below 10 cm/s). In addition, the ratio
between the largest and the shortest times is found to be very close to 2, for
both sizes. Finally, the velocity above which a plateau is observed increases
with the drop size.

All these observations are qualitatively in agreement with our predictions. A
fully quantitative comparison would require a very accurate measurement of the
drop radii (which determine the value of $V_{c}$), and also taking into
account the possibility for the drop to stick at very small velocity: for
$V\sim V_{c}$, it is observed that the drop does not bounce off but remains
stuck to the solid. In contrast, the comparison between the model and the data
becomes quite precise at higher speeds: the plateau value is well described by
Eq. (\ref{4}) with a numerical coefficient of 2.6$\pm0.1$ \cite{DR}, in good
agreement with 2.3, the value obtained above.

\begin{figure}[!]
\twofigures[scale=0.3]{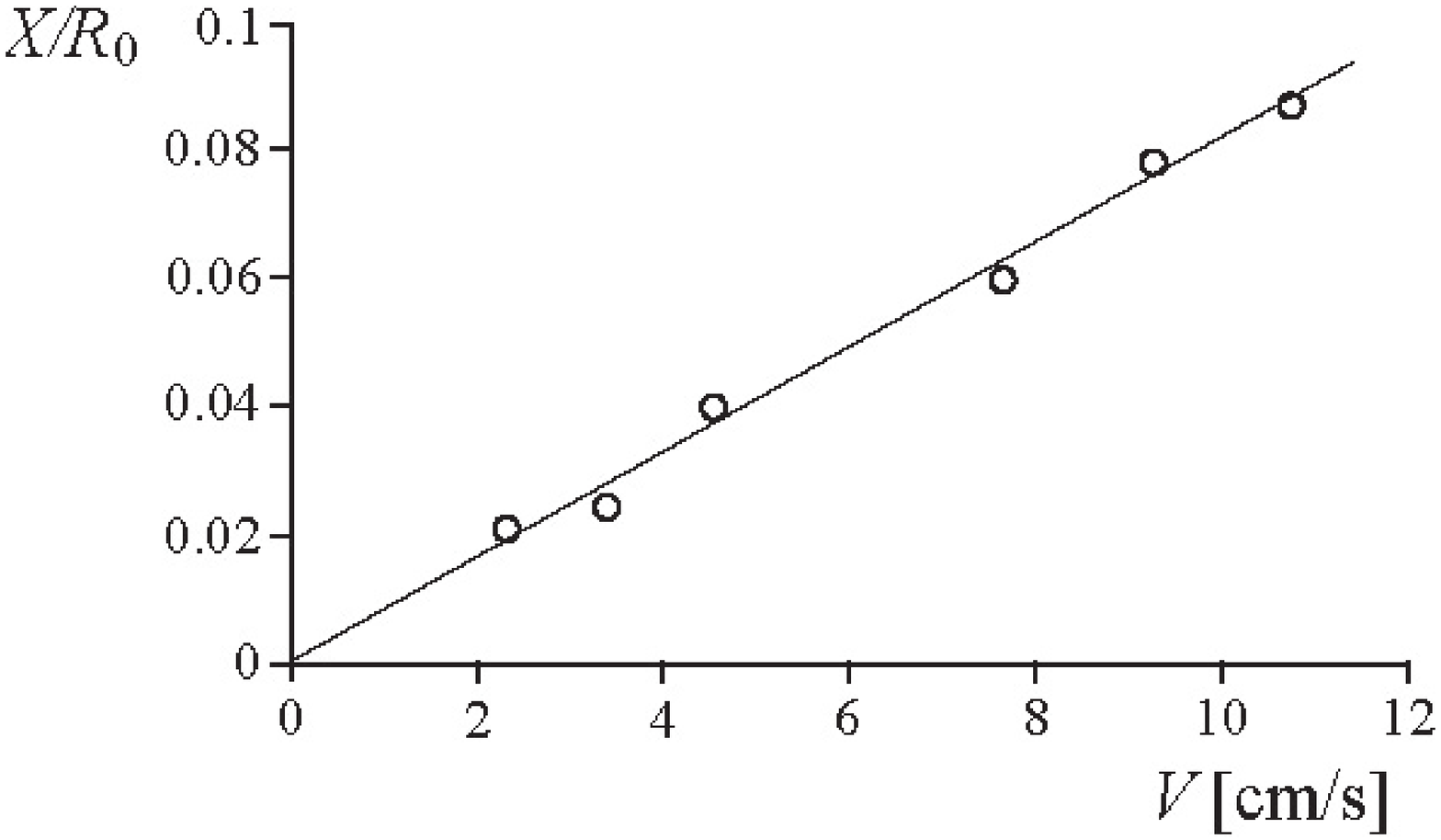}{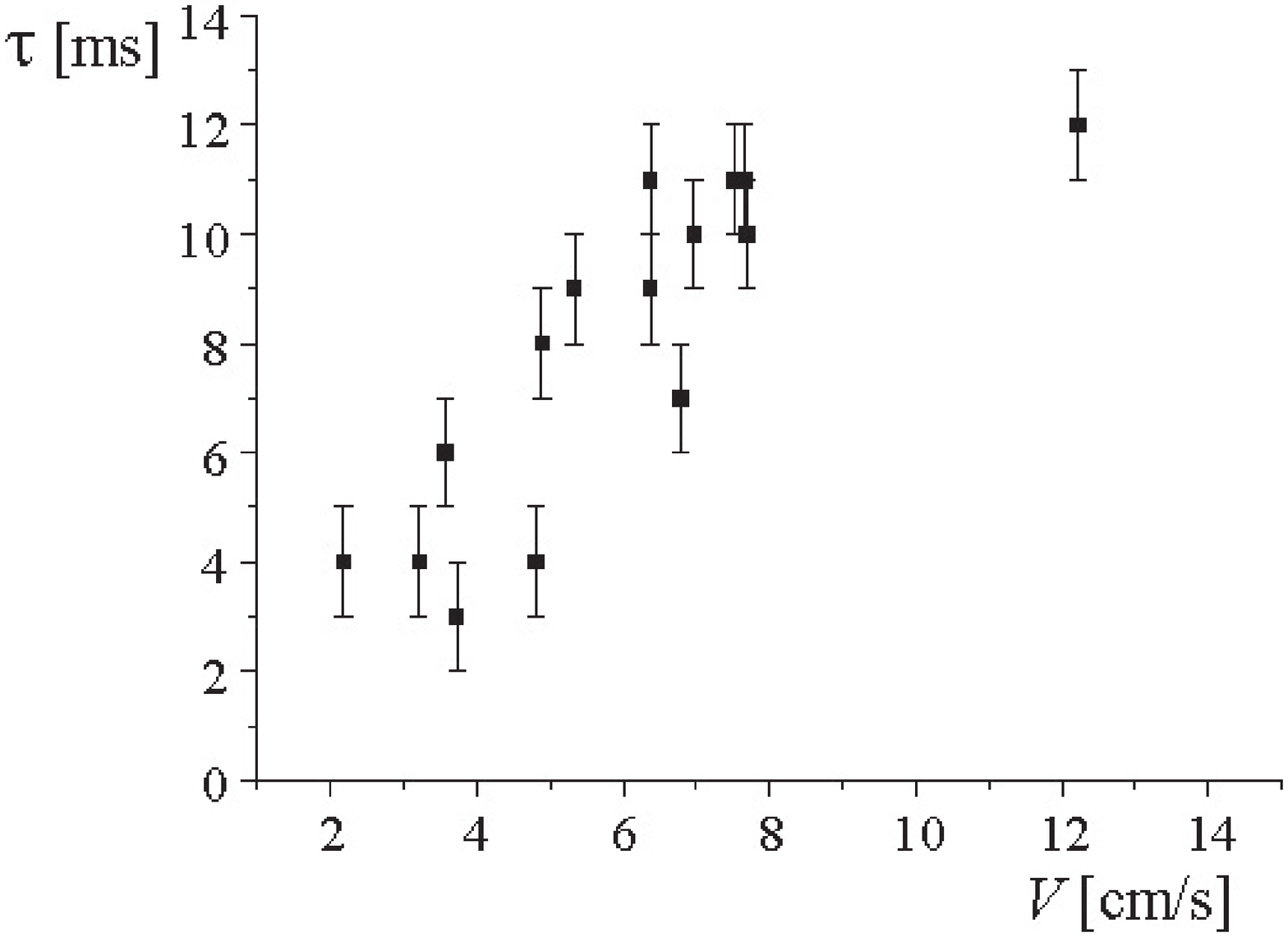}
\caption{Maximal deformation $X$ of a water drop ($R_{0}\sim0.4$
mm) as a function of its impact velocity. Experimental data
(circles) are compared with the scaling prediction (line) of Eq. (6).}
\label{f5}
\caption{Contact time of a water drop ($R_{0}\sim1$mm)
in the field of an inverted gravity.}
\label{f6}
\end{figure}

We also measured the maximal deformation $X$ of the drop during the impact as
a function of the impact velocity $V$ (Fig. \ref{f5}).
We found a linear behavior, in agreement with Eq. (\ref{3}).

This first series of experiments confirms that the contact time of the
bouncing drop deviates from its asymptotic value when we are in the regime of
low impact velocity, \textit{i.e.} in the linear regime of deformation. We
interpreted this effect as due to gravity, and tried to confirm this
interpretation thanks to a second series of experiments. There, we studied the
bouncing with gravity working in the opposite direction (\textit{inverted
gravity}). Drops were dropped from centimetric heights onto a
super-hydrophobic substrate slightly tilted. After bouncing off, they hit at a
height close to their maximum rebound height on a second plate of the same
nature and inclined by the same angle. We recorded the contact time for
rebounds on this second plate. The data are plotted in Fig. \ref{f6}, as a
function of the impact velocity. The error bars are there larger (because of
the experimental resolution and the effects due to the vibrations induced by
the first impact), but the data clearly indicate that the contact time
\textit{increases} with the impact velocity. This is in agreement with our
theory: transforming $g$ into $-g$ (equivalently, $\delta$ into $-\delta$) in
Eq. (\ref{analytical}) makes $\tau$ increase with $V$ as plotted in dotted
line in Fig. \ref{f3}. Interestingly, here, the contact time can vary between
$0$ and $\tau$ (between $\tau$ and 2$\tau$ in the previous experiments) which
can be understood qualitatively using the construction suggested in Fig.
\ref{f2}. In the inverted gravity case, the whole sequence takes place below
the line $x=0$, so that we recover that the contact time should vary between
$0$ and $\tau$.

\section{Perspectives}

Interesting perspectives can be given in the regime of higher impact
velocities ($V>V_{W}$) where the drop is highly deformed and makes some kind
of (transient) pancake of radius $R_{0}+X\sim X$ and thickness $h$ (Fig.
\ref{f1}.b). Since the drop crashes on the solid during a \textit{crashing
time} $R_{0}/V$, we expect the inertial term in the Euler's equation to be of
the order of $\rho V^{2}/R_{0}$. The Laplace pressure scales as $\gamma/h$ at
the equator and it changes over the length $h$. The pressure gradient thus
scales as $\gamma/h^{2}$. The gravity term can be neglected in this regime of
large deformation, so that the Euler's equation can be cast dimensionally into%
\begin{equation}
\rho R_{0}^{3}V^{2}\sim\gamma X^{4}/R_{0}^{2} \label{enrgcons2}%
\end{equation}
\textit{via} the conservation of the volume ($R_{0}^{3}\sim hR^{2}$). Thus,
the maximal size of the pancake is deduced as%
\begin{equation}
X\sim R_{0}W^{1/4} \label{addlarge}%
\end{equation}
where $W$ is defined in Eq. (\ref{weber}). $X$ is indeed found to be larger
than $R_{0}$ for $W>1$. We note that if the initial kinetic energy were
transformed mainly into a surface term, namely,
\begin{equation}
\rho R_{0}^{3}V^{2}\sim\gamma X^{2}, \label{hypo}%
\end{equation}
we would find $X\sim R_{0}W^{1/2}$ instead of Eq. (\ref{addlarge}).\ 

Eq. (\ref{enrgcons2}) is in striking contrast with the case of a small
deformation where the kinetic energy was found to be stored in surface (and
gravitational) energy during the shock as in Eq. (\ref{ec2}). Namely, Eq.
(\ref{enrgcons2}) tells us that the initial energy is transferred primarily
into other forms, identities of which (internal flow in the pancake, and
possibly, the phonon energy of the substrate, etc.) remains to be clarified.

In this regime, the contact time can be regarded as the dewetting time of the
pancake. In this inertial case, the dewetting velocity $V_{d}$ scales as
$\sqrt{\gamma/\left(  \rho h\right)  }$ \cite{text}, which corresponds to the
retraction speed of a liquid sheet \cite{Taylor}. Thus, the contact time is
given by $\tau\sim X/V_{d}$, which happens to result in Eq. (\ref{4}). Indeed,
the contact time was observed to be independent of $V$ and to increase as
$R_{0}^{3/2}$ even for such high velocities \cite{DR}. This fact emphasizes
that the drop is no longer in a linear-spring regime, for which Eq.
(\ref{hypo}) would hold, and the contact time would be given by $\tau\sim
X/V$. Note also that both $X/V_{d}$ and $X/V$ are certainly much longer than
the crashing time $R_{0}/V$ in the present case ($W\gg1$), which justifies our
previous estimation of the inertial term in Eq. (\ref{enrgcons2}).

\section{Conclusion}

In the limit of small deformations (\textit{i.e.} for small impact
velocities), a liquid ball thrown on a solid behaves as a quasi-ideal spring.
This can be understood as a conventional spring-mass system with a stiffness
given by surface tension and a mass given by that of the ball; the deformation
of the small ball during the impact linearly depends on the impact velocity
and the contact time scales as the period of this spring-mass system, as
observed with high speed photographs. The contact time was found to
\emph{increase} (typically by a factor of 2) at small impact velocity, which
can be interpreted as the result of the weight of the ball. This was confirmed
by achieving a similar experiment in an inverted field of gravity, which
indeed leads to a \emph{decrease} of the contact time at small velocities.

In this system, the effect of viscosity could be ignored. This might be
physically due to the absence of a contact line in a situation of non-wetting;
most of viscous dissipation usually takes place near the contact. It would be
interesting to see how our views would be modified when the viscosity of the
liquid consisting the ball is increased. Extensions to similar (but different)
systems such as gel balls \cite{Tanaka} or drops of surfactant solution would
also be worth studying.

\begin{acknowledgements}%
We thank Yoshimi Tanaka for useful discussions.
K. O. is grateful to P.-G. de Gennes and members of his group
for warm hospitality during his second and third stay
in Paris. The second stay was financially supported by
Joint Research Project between JSPS and CNRS while
the third by Coll\`{e}ge de France.
This work is also supported by an internal grant of Ochanomizu University.
\end{acknowledgements}


\end{document}